\documentclass[runningheads]{llncs}
\usepackage{graphicx}
\usepackage{comment}
\usepackage{amsmath,amssymb} 
\usepackage{color}
\usepackage{subcaption}
\usepackage{booktabs}

\begin{document}
\pagestyle{headings}
\mainmatter
\def\ECCVSubNumber{100}  

\title{UDC 2020 Challenge on Image Restoration of Under-Display Camera: Methods and Results} 

\newcommand{\ff}[1]{\textcolor{blue}{#1}}

\titlerunning{Under-Display Camera Challenge: Methods and Results}
%
\author{Yuqian Zhou, Michael Kwan, Kyle Tolentino, 
Neil Emerton, Sehoon Lim, Tim Large, Lijiang Fu, Zhihong Pan, Baopu Li, Qirui Yang, Yihao Liu, Jigang Tang, Tao Ku, Shibin Ma, Bingnan Hu, Jiarong Wang, Densen Puthussery, Hrishikesh P S, Melvin Kuriakose, Jiji C V, Varun Sundar, Sumanth Hegde, Divya Kothandaraman, Kaushik Mitra, Akashdeep Jassal, Nisarg A. Shah, Sabari Nathan, Nagat Abdalla Esiad Rahel, Dafan Chen, Shichao Nie, Shuting Yin, Chengconghui Ma, Haoran Wang, Tongtong Zhao, Shanshan Zhao, Joshua Rego, Huaijin Chen, Shuai Li, Zhenhua Hu, Kin Wai Lau, Lai-Man Po, Dahai Yu, Yasar Abbas Ur Rehman,Yiqun Li, Lianping Xing}
\authorrunning{Zhou et al.}
%
\institute{}
\maketitle

\begin{abstract}
This paper is the report of the first Under-Display Camera (UDC) image restoration challenge in conjunction with the RLQ workshop at ECCV 2020. The challenge is based on a newly-collected database of Under-Display Camera. The challenge tracks correspond to two types of display: a 4k Transparent OLED (T-OLED) and a phone Pentile OLED (P-OLED). Along with about 150 teams registered the challenge, eight and nine teams submitted the results during the testing phase for each track. The results in the paper are state-of-the-art restoration performance of Under-Display Camera Restoration. Datasets and paper are available at https://yzhouas.github.io/projects/UDC/udc.html.

\end{abstract}

\section{Introduction}
Under-Display Camera (UDC)~\cite{zhou2020image} is specifically designed for full-screen devices as a new product trend, eliminating the need for bezels. Improving the screen-to-body ratio will enhance the interaction between users and devices. Mounting the display in front of a camera imaging lens will cause severe image degradation like low-light and blur. It is then desirable to propose proper image restoration algorithms for a better imaging quality identical to the original lens. It will also potentially benefit the downstream applications like object detection\cite{yu2020scale} and face analysis\cite{zhou2018survey}. 

Obtaining such algorithms can be challenging. First, most existing methods leverage the advantages of deep learning to resolve multiple image degradation problems, such as image denoising\cite{abdelhamed2019ntire,abdelhamed2020ntire,zhou2019awgn,zhou2019adaptation,liu2019learning}, deblurring\cite{guo2019effects}, super-resolution\cite{mei2020image,mei2020pyramid} etc. Restoration of UDC images, as a problem of recovering combined degradation, requires the joint modeling of methods resolving different optical effects caused by the displays and camera lens. It also requires the researchers to understand inter-disciplinary knowledge of optics and vision. Second, data acquisition process can be challenging due to variant types of displays and cameras. Collecting data sets consisting of pairs of degraded and undegraded images that are in other respects identical is challenging even using special display-camera combination hardware. Furthermore, the trained model may not be easily generalized to other devices.. 

In this paper, we report the methods and results from the participants of the first Under-Display Camera Challenge in conjunction with the Real-world Recognition from Low-quality Inputs (RLQ) workshop of ECCV 2020. We held this image restoration challenge to seek an efficient and high-performance image restoration algorithm to be used for recovering under-display camera images. Participants greatly improved the restoration performance compared to the baseline paper. More details will be discussed in the following sections. 

\section{Under-Display Camera (UDC) Challenge}
\subsection{UDC Dataset}
The UDC dataset is collected using a monitor-based imaging system as illustrated in the baseline paper\cite{zhou2020image}. Totally 300 images from DIV2K\cite{agustsson2017ntire} dataset are displayed on the monitor screen, and paired data is recorded using a FLIR Flea camera. In this challenge, we only use the 3-channel RGB data for training and validation. The training data consists of 240 pairs of $1024\times2048$ images, totally 480 images. Validation and Testing inputs each consist of 30 images of the same resolution. The challenge is organized in two phases: validation and testing. We only release the ground truth of the validation set after the end of the validation phase, while the ground truth of the testing partition is kept hidden.

\subsection{Challenge Tracks and Evaluation}
The challenge had two tracks: \textbf{T-OLED} and \textbf{P-OLED} image restoration. Participants were encouraged to submit results on both of them, but only attending one track was also acceptable. For both tracks, we evaluated and ranked the algorithms using the standard Peak Signal To Noise Ratio (PSNR). Additional measurements like Structural Similarity (SSIM) and inference time are also reported for reference. Although an algorithm with high efficiency is extremely important for portable devices, we did not rank the participants based on the inference time. In total of 83 teams took part in the T-OLED track, and 73 teams registered the P-OLED track. Finally, 8 teams submitted the testing results to T-OLED track, and 9 teams to the P-OLED track.

\section{Teams and Methods}
In this section, we summarize all the methods from the participants who submitted the final results and reports for each track. 
\subsection{Baidu Research Vision}
\begin{figure}[t!]
  \centering
  \includegraphics[scale=0.45]{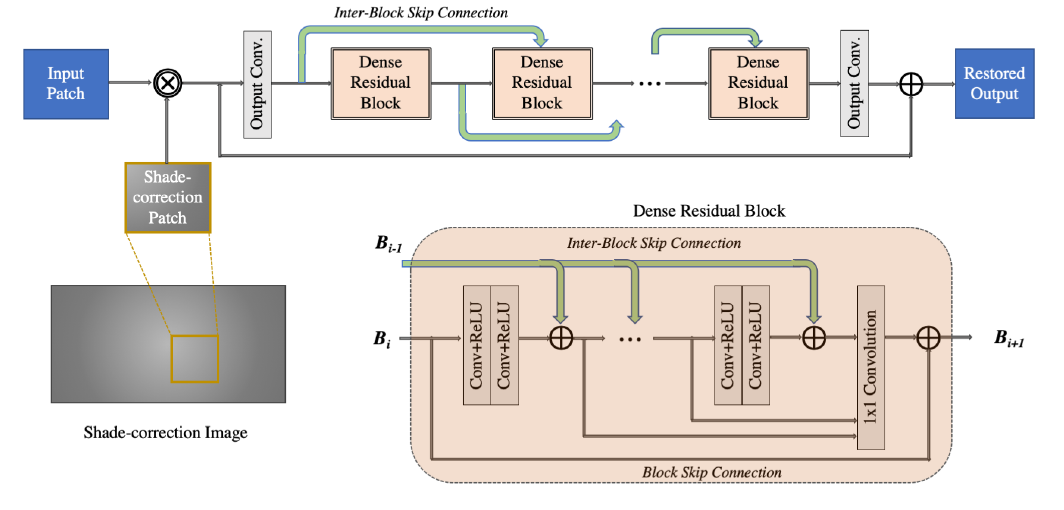}
  \caption{The Dense Residual Network architecture proposed by Team Baidu Research Vision.}
  \label{fig:Baidu}
\end{figure}

\textbf{Members:} \textbf{Zhihong Pan}, Baopu Li\\
\textbf{Affiliations:} Baidu Research (USA)\\
\textbf{Track}: T-OLED\\
\textbf{Title: Dense Residual Network with Shade-Correction for UDC Image Restoration} The architecture proposed by Team Baidu Research Vision is shown in Fig. \ref{fig:Baidu}. The team branches off of prior work of a dense residual network for raw image denoising and demosaicking \cite{abdelhamed2020ntire} with a newly added novel shade-correction module. The novel shade-correction module consists of a set of learned correction-coefficients with the same dimension as the full-size image. Matching coefficients of the input patch are multiplied with the input for shade correction. The proposed shade-correction module could learn patterns related to the specific T-OLED screen, so additional learning might be needed for different set-ups.
However, the team believes this fine-tuning process tends to be efficient. The model is trained on patches of size $128\times128$.

\subsection{BigGuy}
\begin{figure}[t!]
\centering
\begin{subfigure}{\textwidth}
  \centering
  \includegraphics[width=1\linewidth]{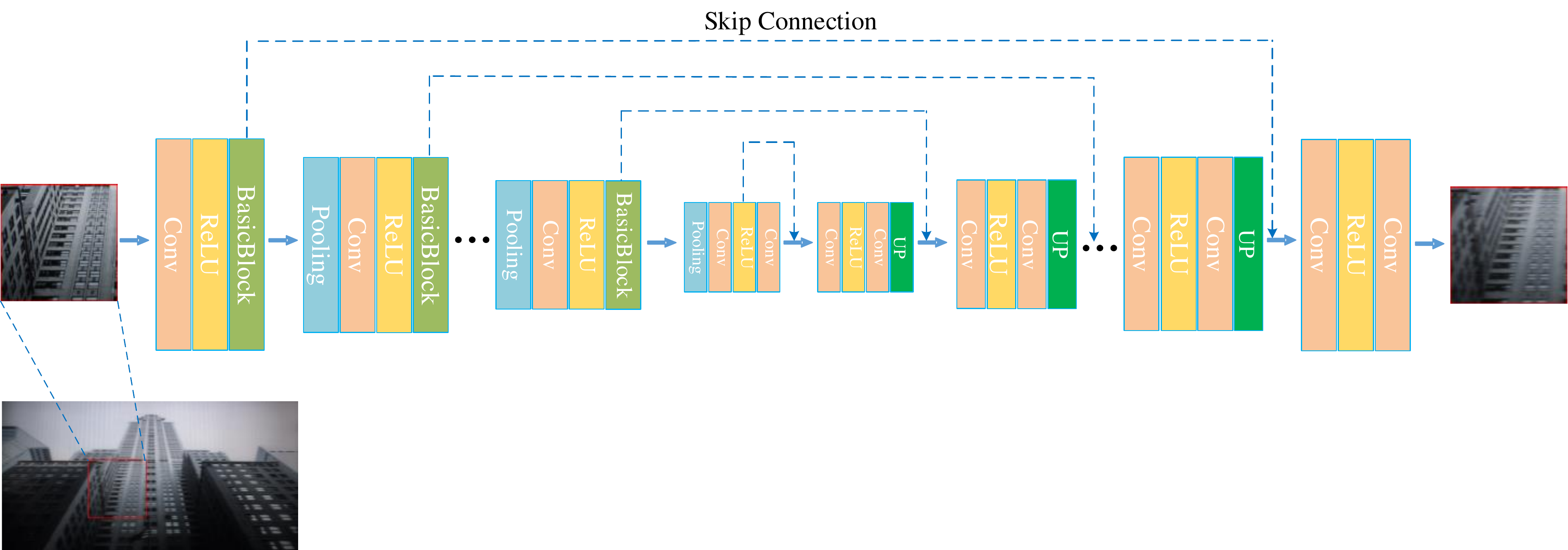}
  \caption{The overall architecture is similar with UNet. We can choose or design different "basic blocks" (e.g. residual block,dense block, residual dense block) to obtain better performance.}
\end{subfigure}
\begin{subfigure}{\textwidth}
  \centering
  \includegraphics[width=0.6\linewidth]{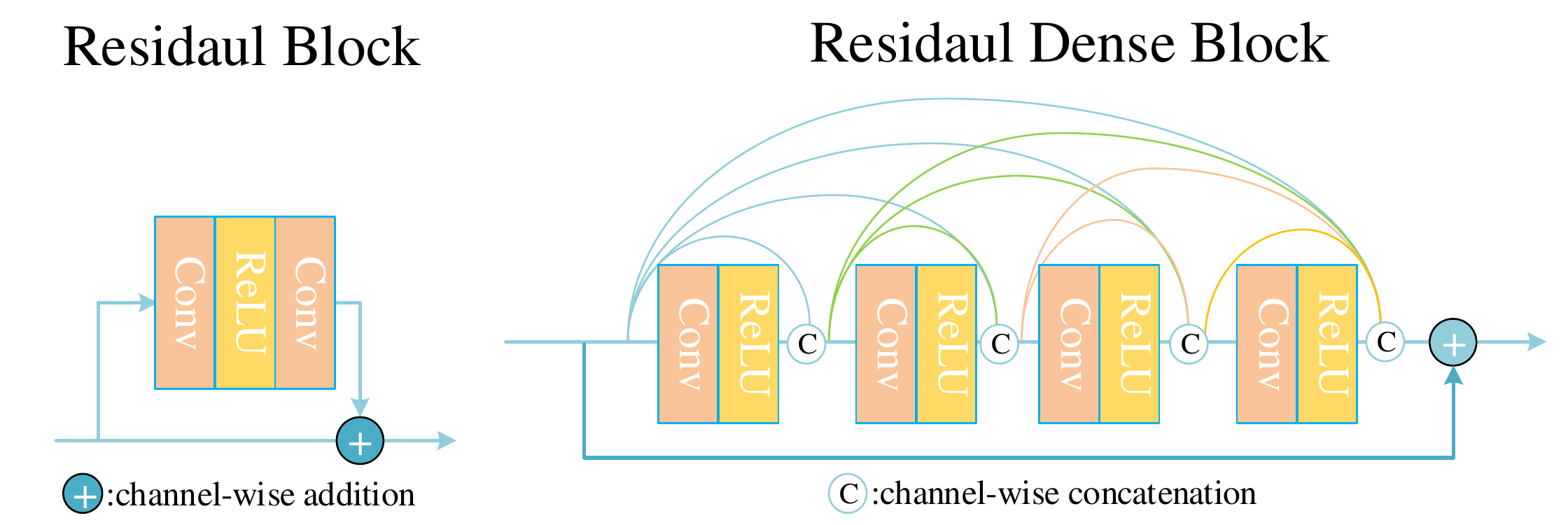}
  \caption{Left: residual block. Right: the proposed residual dense block. It consists of a residual connection and a dense module composed of four convolutional layers.
}
\end{subfigure}
\caption{The architectures proposed by Team BigGuy. }
\label{fig:BigGuy}
\end{figure}

\textbf{Members:} \textbf{Qirui Yang}, Yihao Liu, Jigang Tang, Tao Ku\\
\textbf{Affiliations:} Chinese Academy of Sciences, Shenzhen Institutes of Advanced Technology\\
\textbf{Track}: T-OLED and P-OLED\\
\textbf{Title: Residual and Dense U-Net\cite{ronneberger2015u} for Under-display Camera Restoration}
The team's method is based on the U-net network. The U-Net encoder network consists of several residual dense blocks. A decoder network is constructed by skip connection and pixel-shuffle upsampling\cite{shi2016real}. Their experiments show that T-OLED and P-OLED have different adaptability to the model and the patch size during training. Therefore, they proposed two different U-Net structures. For the T-OLED track, they
used Residual Dense Blocks as the basic structure and proposed a Residual-Dense U-Net model (RDU-Net) as shown in Fig. \ref{fig:BigGuy}. 

For the P-OLED track, they found P-OLED panels allow small amounts of light to enter the camera so that P-OLED images present a dull feature. The difference between input and output images of P-OLED dataset is mainly reflected in color and high-frequency texture information. They thus explored residual dense blocks~\cite{zhang2020residual}, ResNet\cite{he2016deep}, and RCAB\cite{zhang2018image} and found residual block achieved the best validation PSNR. The model structure for the P-OLED track is shown in Fig.\ref{fig:BigGuy}.

\subsection{BlackSmithM}
\textbf{Members:} \textbf{Shibin Ma}, Bingnan Hu, Jiarong Wang\\
\textbf{Affiliations:} None\\
\textbf{Track}: P-OLED\\
\textbf{Title: P-OLED Image Reconstruction Based on GAN Method} In view of the poor quality blurred image of the P-OLED track, the team focused on adjusting the light first. After the image is dimmed, they can better remove the blur. The team used the pix2pix \cite{isola2017image,zhu2017unpaired} model to adjust the poor image light, and at the same time, it was possible to recover the image information. Before the image passed the pix2pix model, the team preprocessed the single scale Retinex (SSR) \cite{wei2018deep}, and croped the $1024\time2048$ data set to the left and right images of $1024\times1024$. The image after the pix2pix network contained noise, so a Gaussian filter was used to process the image to make the resulting image more smooth and real, thus improving the PSNR value of the image. 

\subsection{CET$\_$CVLab}
\begin{figure}[t!]
\centering
\begin{subfigure}{\textwidth}
  \centering
  \includegraphics[width=1\linewidth]{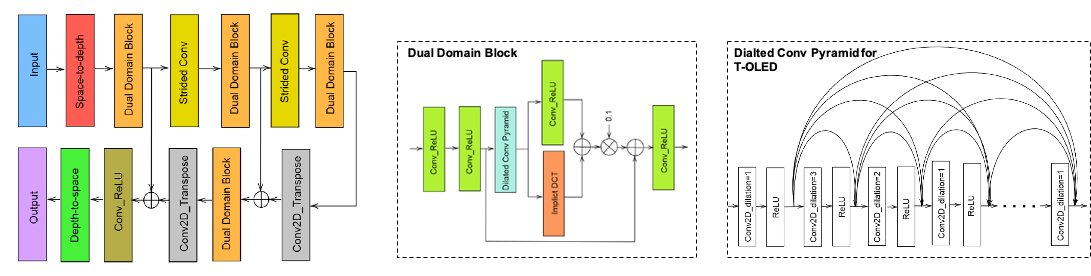}
  \caption{The Dual Domain Net for T-OLED Track.}
\end{subfigure}
\begin{subfigure}{\textwidth}
  \centering
  \includegraphics[width=1\linewidth]{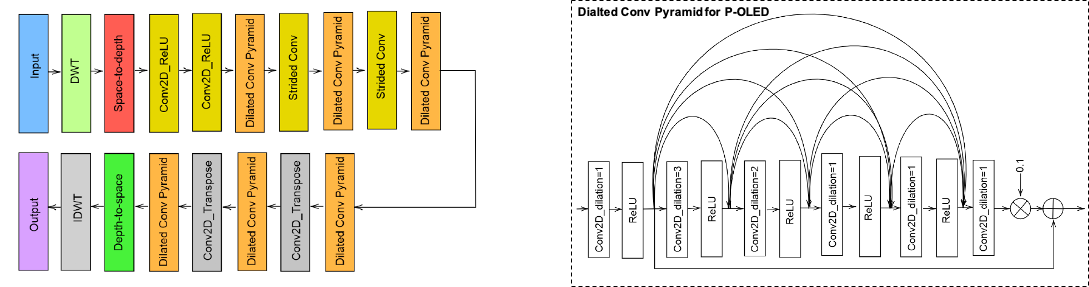}
  \caption{The PDCRN architecture for P-OLED Track }
\end{subfigure}
\caption{The architectures proposed by Team CET$\_$CVLab. }
  \label{fig:CET CVLab}
\end{figure}

\textbf{Members:} \textbf{Densen Puthussery}, Hrishikesh P S, Melvin Kuriakose, Jiji C V\\
\textbf{Affiliations:} College of Engineering, Trivandrum, India \\
\textbf{Track}: T-OLED and P-OLED\\
\textbf{Title: Dual Domain Net (T-OLED) and Wavelet decomposed dilated pyramidal CNN (P-OLED)} The team proposed encoder-decoder structures to learn the restoration. For the T-OLED track, they proposed a dual-domain net (DDN) inspired by \cite{zheng2019implicit}. In the dual domain method, the image features are processed in both pixel domain and frequency domain using implicit discrete cosine transform. This enables the network to correct the image degradation in both frequency and pixel domain, thereby enhancing the restoration. The DDN architecture is shown in Fig. \ref{fig:CET CVLab}(a).

For the P-OLED track, inspired by the multi-level wavelet-CNN (MWCNN) proposed by
Liu et al. \cite{liu2018multi}, they proposed Pyramidal Dilated Convolutional RestoreNet (PDCRN) which follows an encoder-decoder structure as shown in Fig. \ref{fig:CET CVLab}(b). In the proposed network, the downsampling operation in the encoder is discrete wavelet transform (DWT) based decomposition instead of down-sampling convolution or pooling. Similarly, in the decoder network, inverse discrete wavelet transform (IDWT) is used instead of upsampling convolution. In the wavelet based decomposition used here, the information from all the channels are combined in the downsampling process to minimize information loss when compared to that of convolutional downsampling. The feature representation for both tracks is made efficient using a pyramidal dense dilated convolutional block. The dilation rate is gradually decreased as the dilated convolution pyramid is moved up. This is to compensate for information loss that may occur with highly dilated convolution due to a non-overlapping moving window in the convolution process.

\subsection{CILab IITM}
\begin{figure}[t!]
  \centering
  \includegraphics[width=\linewidth]{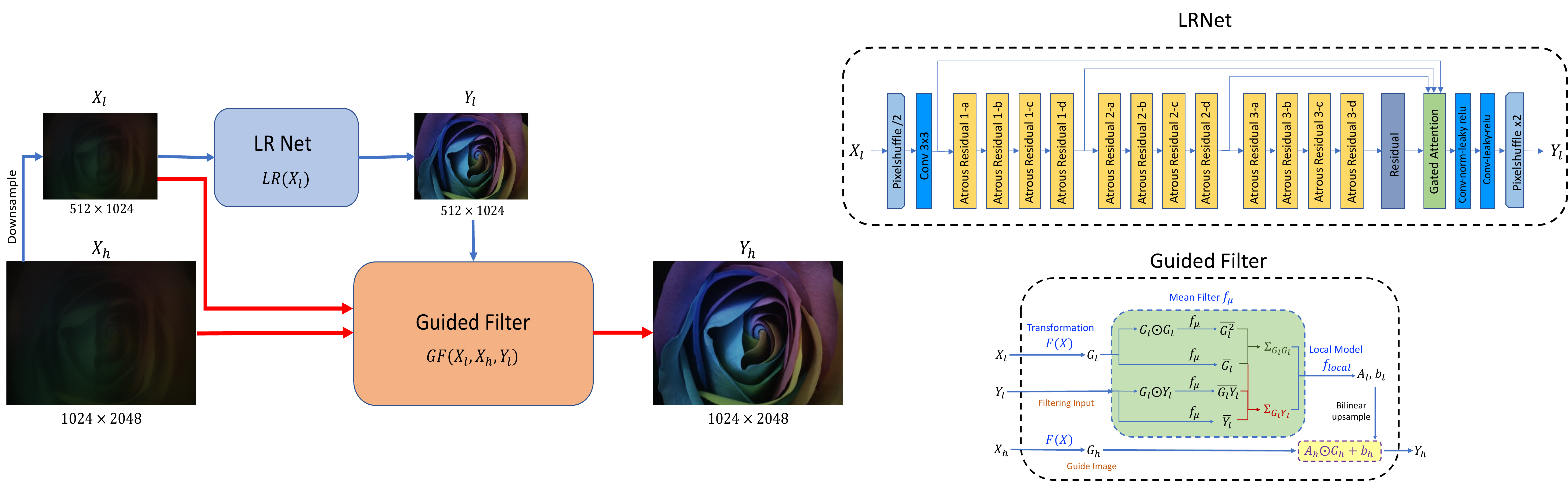}
  \caption{The Deep Atrous Guided Filter architectures of the LRNet and the guided filter proposed by Team CILab IITM.}
  \label{fig:cilab_iitm}
\end{figure}

\textbf{Members:} \textbf{Varun Sundar}, Sumanth Hegde, Divya Kothandaraman, Kaushik Mitra\\
\textbf{Affiliations:} Indian Institute of Technology Madras \\
\textbf{Track}: T-OLED and P-OLED\\
\textbf{Title: Deep Atrous Guided Filter for Image Restoration in Under Display
Cameras.} The team uses two-stage pipeline for the task as shown in Fig. \ref{fig:cilab_iitm}. The first stage is a
low-resolution network (LRNet) which restores image quality at low-resolution. The low resolution network retains spatial resolution and emulates multi-scale information fusion with multiple atrous convolution blocks \cite{brehm2020high,chen2019gated} stacked in parallel. In the second stage, they leverage a guided filter to produce a high resolution image from the low resolution refined image obtained from stage one. They further propose a simulation scheme to augment data and boost performance. More details are in the team's challenge report\cite{sundar2020deep}.

\subsection{Hertz}
\textbf{Members:} \textbf{Akashdeep Jassal$^{1}$}, Nisarg A. Shah$^{2}$\\
\textbf{Affiliations:} Punjab Engineering College, India$^1$, IIT Jodhpur$^{2}$ \\
\textbf{Track}: P-OLED\\
\noindent\textbf{Title: P-OLED reconstruction using GridDehazeNet} Based on GridDehazeNet which aims to clear haze from a low resolution image \cite{liu2019griddehazenet}, the team uses the network which contains a pre-processing residual dense block, a grid-like backbone of the same residual dense blocks interconnected with convolutional downsampling and upsampling modules, and a post-processing stage of another residual dense block.

\subsection{Image Lab}
\begin{figure}[t!]
  \centering
  \includegraphics[width=\linewidth]{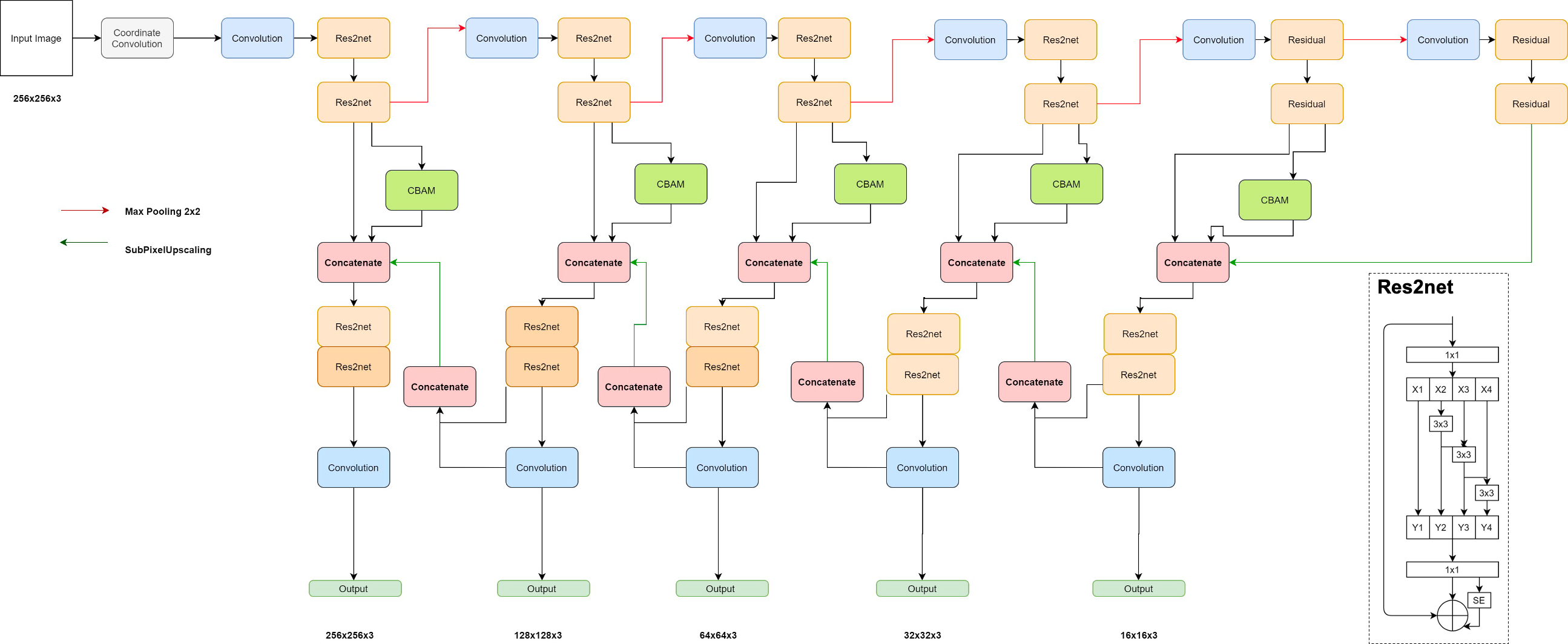}
  \caption{The Lightweight Multi Level Supervision Model architecture proposed by Team Image Lab.}
  \label{fig:Image Lab}
\end{figure}

\textbf{Members:} \textbf{Sabari Nathan$^{1}$}, Nagat Abdalla Esiad Rahel$^{2}$\\
\textbf{Affiliations:} Couger Inc, Japan$^{1}$, Al-Zintan University, Libya$^{2}$ \\
\textbf{Track}: T-OLED and P-OLED\\
\noindent\textbf{Title: Image Restoration using Light weight Multi Level Supervision
Model} The team proposes a Lightweight Multi Level Supervision Model inspired by \cite{nathan2020moire}. The architecture is shown in Fig. \ref{fig:Image Lab}. The input image is first passed to the coordinate convolutional layer to map the pixels to a Cartesian coordinate space\cite{liu2018intriguing}, and then fed into the encoder. The encoder composed of $3\times3$ convolution layers, two Res2Net\cite{gao2019res2net} blocks, and a downsampling layer, while the decoder block replaces the last component with a subpixel scaling layer\cite{shi2016real}. A convolution block attention module (CBAM)\cite{woo2018cbam} in the skip connection is concatenated with the encoding block as well.

\subsection{IPIUer}
\begin{figure}[t!]
  \centering
  \includegraphics[scale=0.1]{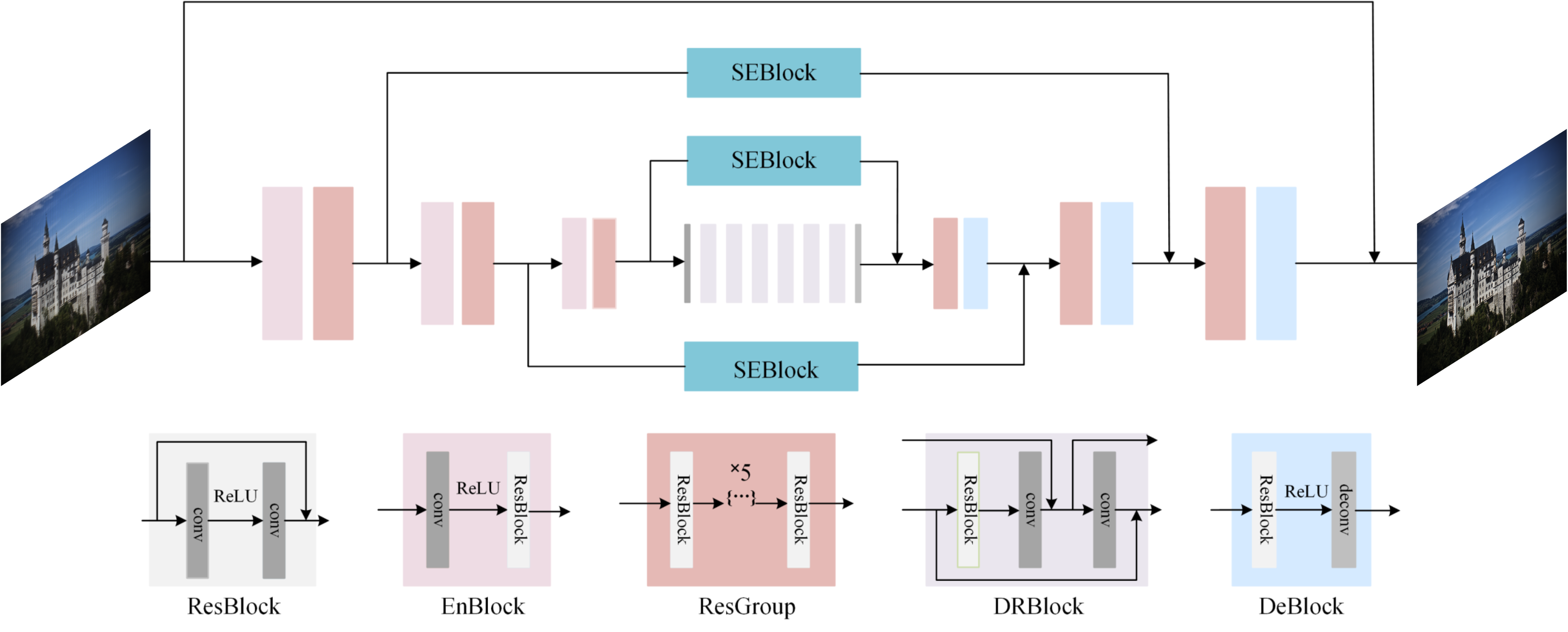}
  \caption{The novel Unet structure proposed by Team IPIUer.}
  \label{fig:IPIUer}
\end{figure}

\textbf{Members:} \textbf{Dafan Chen}, Shichao Nie, Shuting Yin, Chengconghui Ma, Haoran Wang\\
\textbf{Affiliations:} Xidian University \\
\textbf{Track}: T-OLED\\
\noindent\textbf{Title: Channel Attention Image Restoration Networks with Dual Residual Connection} The team proposes a novel UNet model inspired by Dual Residual Networks\cite{liu2019dual} and Scale-recurrent
Networks (SRN-DeblurNet)\cite{tao2018scale}. As shown in Fig. \ref{fig:IPIUer}, in the network, there are 3 EnBlocks, 3 DeBlocks and 6 DRBlocks. The
entire network has three stages and one bridge between encoder and decoder. Every stage consists of one EnBlock/DeBlock and a residual group (ResGroup). The ResGroup has seven residual blocks (ResBlock). Between the encoder and decoder, a dual residual block (DRBlock) is used to extract high level semantic features effectively.. The skip connection uses squeeze-and-exitation blocks \cite{hu2018squeeze} which aims to highlight the features of some dimensions. 

\subsection{lyl}
\begin{figure}[t!]
  \centering
  \includegraphics[width=0.6\linewidth]{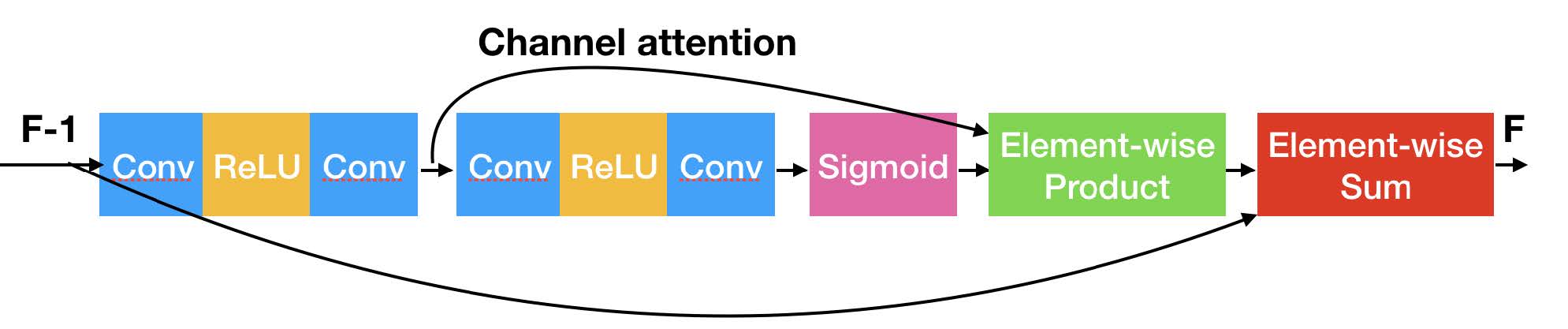}
  \caption{The Lightweight Upsampling Module (LUM) named FineNet proposed by Team lyl.}
  \label{fig:lyl}
\end{figure}

\textbf{Members:} \textbf{Tongtong Zhao}, Shanshan Zhao\\
\textbf{Affiliations:} Dalian Maritime University \\
\textbf{Track}: T-OLED and P-OLED\\
\noindent\textbf{Title: Coarse to Fine Pyramid Networks for Progressive Image Restoration} The team proposes a coarse to fine network (CFN) for progressive reconstruction. Specifically, in each network level, the team proposes a lightweight upsampling module (LUM), also named FineNet as in Fig. \ref{fig:lyl}, to process the input, and merge it with the input features. Such progressive cause-and-effect process helps
to achieve the principle for image restoration: high-level information can guide an image to recover a better restored image. The authors claim that they can achieve competitive results with a modest number of parameters. 

\subsection{San Jose Earthquakes}
\begin{figure}[t!]
  \centering
  \includegraphics[scale=0.08]{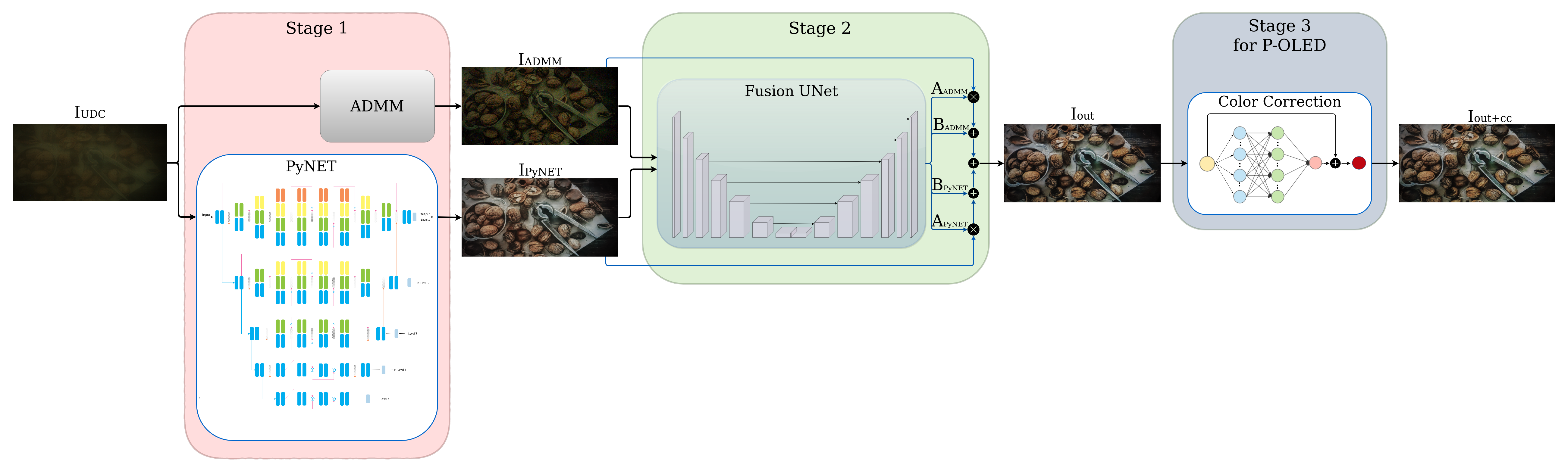}
  \caption{The network architecture of multi-stage restoration proposed by Team San Jose Earthquakes.}
  \label{fig:San Jose Earthquakes}
\end{figure}

\textbf{Members:} \textbf{Joshua Rego}, Huaijin Chen, Shuai Li, Zhenhua Hu\\
\textbf{Affiliations:} SenseBrain Technology \\
\textbf{Track}: T-OLED and P-OLED\\
\noindent\textbf{Title: Multi-stage Network for Under-display Camera Image Restoration} The team proposes multiple stage networks as shown in Fig. \ref{fig:San Jose Earthquakes} to solve different issues caused by under-display cameras. For T-OLED, the pipeline is two-staged. The first stage uses a multi-scale network PyNET\cite{ignatov2020replacing} to recover intensity and largely apply deblur to the input image. The second stage is a U-Net fusion network that uses the output of the PyNET as well as a sharper, but noisier, alternating direction method of multipliers (ADMM)\cite{boyd2011distributed} reconstruction as the inputs of the network and outputs weights used to combine the two inputs for a sharper, de-noised result. P-OLED pipeline uses an additional third stage color correction network to improve color consistency with the target image by pixel-wise residual learning. 

The authors mentioned that training solely through the first-stage network, while greatly restoring towards the target image, was unable to restore sharpness completely, especially in higher frequency textures. The ADMM output, on the other hand, was able to retain high-frequency sharpness, but suffered largely from noise and some additional artifacts. The fusion network blends the desirable characteristics from the two to slightly improve the results. However, one drawback of the method is that ADMM takes about 2.5 mins to solve each image, which is longer for inference. Nevertheless, the method is a novel approach fusing the implementation of a traditional
and deep-learning method.

\subsection{TCL Research HK}
\begin{figure}[t!]
  \centering
  \includegraphics[width=0.6\linewidth]{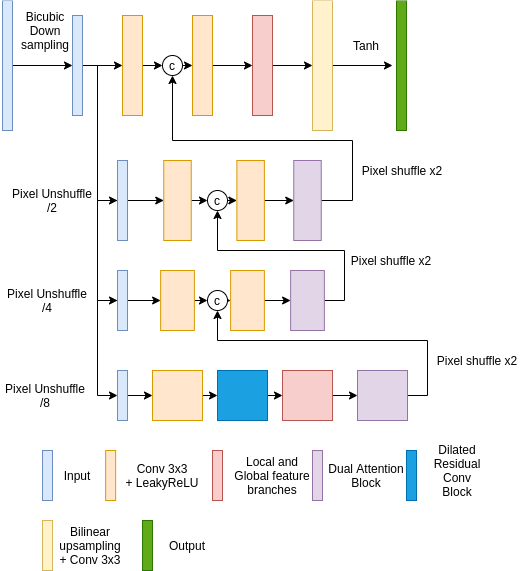}
  \caption{The network architecture proposed by Team TCL Research HK.}
  \label{fig:TCL Research HK}
\end{figure}

\textbf{Members:} \textbf{Kin Wai Lau$^{1,2}$}, Lai-Man Po$^{1}$, Dahai Yu$^{2}$, Yasar Abbas Ur Rehman$^{1,2}$, Yiqun Li$^{1,2}$, Lianping Xing$^{2}$\\
\textbf{Affiliations:} City University of Hong Kong $^{1}$, TCL Research Hong Kong $^{2}$ \\
\textbf{Track}: P-OLED\\
\noindent\textbf{Title: Self-Guided Dual Attention Network for Under Display Camera Image
Restoration} The team proposes a multi-scale self-guidance neural architecture containing (1) multi-resolution convolutional branch for extracting multi-scale information, (2) low-resolution to high-resolution feature extraction for guiding the intermediate high-resolution feature extraction process, (3) spatial and channel mechanisms for extracting contextual information, (4) Dilated Residual Block (DRB) to increase the receptive field for preserving the details, (5) local and global feature branch for adjusting local information (e.g., contrast, detail, etc.) and global information (e.g., global intensity, scene category, color distribution, etc.) The network architecture proposed by Team TCL Research HK for Self-Guided Dual Attention Network is shown in Fig. \ref{fig:TCL Research HK}. 

\section{Results}
\begin{table*}[t!]
\newcommand{\len}{9mm}
\newcommand{\cl}{\centering\arraybackslash}
\centering
\resizebox{\textwidth}{!}{
\begin{tabular}
{llcccccccc}
\toprule
{Team} & {Username} & PSNR & SSIM & TT(h) & IT (s/frame)  & {CPU/GPU} & {Platform}  & Ensemble & Loss\\
\midrule
Baidu Research Vision & zhihongp & $38.23_{(1)}$ & $0.9803_{(1)}$ & 12& 11.8 & Tesla M40 &PaddlePaddle, PyTorch & flip($\times 4$) &$L_1$,SSIM \\
IPIUer & TacoT & $38.18_{(2)}$ & $0.9796_{(3)}$ &30  &1.16  &Tesla V100  &PyTorch  &None  &$L_1$\\ 
BigGuy	&JaydenYang	&$38.13_{(3)}$	&$0.9797_{(2)}$&23  &0.3548  & Tesla V100 &PyTorch  & -& $L_1$\\
CET$\_$CVLAB	&	hrishikeshps&	$37.83_{(4)}$	&$0.9783_{(4)}$&72  & 0.42 &Tesla P100  &Tensorflow  &- &$L_2$\\
CILab IITM	&ee15b085	&$36.91_{(5)}$	&$0.9734_{(6)}$&96  &1.72  &GTX 1080 Ti  &PyTorch  &flip($\times 4$) model($\times 8$) &$L_1$ \\
lyl&	tongtong	&$36.72_{(6)}$	&$0.9776_{(5)}$&72  &3  &-  &PyTorch  & model(-)&$L_1$ \\
Image Lab&	sabarinathan&	$34.35_{(7)}$	&$0.9645_{(7)}$&-  &1.6  &GTX 1080 Ti  &Keras &  model(-) &$L_2$,SSIM \\
San Jose Earthquakes&	jdrego	&$33.78_{(8)}$	&$0.9324_{(8)}$&18  &180  &- &PyTorch &model(-)  &- \\
\bottomrule 
\end{tabular}
} 
\caption{Results and rankings of methods of T-OLED Track. TT: Training Time. IT: Inference Time}
 \label{tab:toled}
\end{table*}

\begin{table*}[t!]
\newcommand{\len}{9mm}
\newcommand{\cl}{\centering\arraybackslash}
\centering
\resizebox{\textwidth}{!}{
\begin{tabular}
{llcccccccc}
\toprule
{Team} & {Username} & PSNR & SSIM & TT(h) & IT (s/frame) & {CPU/GPU} & {Platform}  & Ensemble & Loss\\
\midrule
CET$\_$CVLAB&	Densen&	$32.99_{(1)}$&	$0.9578_{(1)}$&72  &0.044  &Tesla T4  &Tensorflow &- &$L_2$ \\
CILab IITM&varun19299&	$32.29_{(2)}$&	$0.9509_{(2)}$&96  &1.72  &GTX 1080 Ti  &PyTorch  &flip($\times 4$) model($\times 8$) &$L_1$\\
BigGuy&	JaydenYang&	$31.39_{(3)}$&	$0.9506_{(3)}$&24&0.2679  &Tesla V100  &PyTorch  &model  & $L_1$\\
TCL Research HK&	stevenlau&	$30.89_{(4)}$&	$0.9469_{(5)}$&48  &1.5&Titan Xp  &PyTorch  & - & $L_1$,SSIM\\
BlackSmithM	&BlackSmithM&	$29.38_{(5)}$&	$0.9249_{(6)}$& - & 2.43& Tesla V100 & PyTorch & - & $L_1$\\
San Jose Earthquakes&	jdrego&	$28.61_{(6)}$&	$0.9489_{(4)}$& 18  &180  &- &PyTorch &model  &- \\
Image Lab&	sabarinathan&	$26.60_{(7)}$&	$0.9161_{(7)}$&- &1.59  & -&-  &None  &$L_2$,SSIM \\
Hertz&	akashdeepjassal&	$25.72_{(8)}$&	$0.9027_{(8)}$& -  &2.29&Tesla K80  &PyTorch  &None  &VGG \\
lyl&	tongtong&	$25.46_{(9)}$&	$0.9015_{(9)}$&72  &3  &-  &PyTorch  & model(-)&$L_1$ \\

\bottomrule 
\end{tabular}
} 
\caption{Results and rankings of methods of P-OLED Track. TT: Training Time. IT: Inference Time}
 \label{tab:poled}
\end{table*}
This section presents the performance comparisons of all the methods in the above sections. The ranking and other evaluation metrics are summarized in Table \ref{tab:toled} for T-OLED track, and in Table \ref{tab:poled} for P-OLED.\\

\textbf{Top Methods}
All the submitted methods are mainly deep-learning oriented. On both tracks, top methods achieved very close PSNR. Among the top-3 methods for T-OLED track, directly training a deep model (e.g. modified UNet) in an end-to-end fashion mostly achieved competitive performance. The outperformed results further demonstrate the effectiveness of using UNet embedding Residual Blocks shared by the teams. Similar structures are widely used in image denoising or deblurring. T-OLED degraded images contain blur and noisy patterns due to diffraction effects, which could be the reason for the superiority of directly applying deblurring/denoising approaches. For the P-OLED track, the winner team, CET$\_$CVLab, proposed to use discrete wavelet transform (DWT) to replace the upsampling and downsampling modules. The CILab IITM team proposed a two-stage pipeline with differentiable guided filter for training. The performance gain can also come from the model pre-trained on the simulated data by using measurements provided by the baseline paper\cite{zhou2020image}. The BigGuy team conducted an extensive model search to find the optimal structures for P-OLED images. Some methods proposed by other teams, though not ranked on top-3, are also novel and worthy to mention. The team San Jose Earthquakes proposed to combine the results of the deep-learning and traditional methods by leveraging the benefits from both ends. The multi-level supervision model proposed by the Image Lab team restores the image in a progressive way. Similarly, the lyl team also share the progressive idea. 

 In addition to module design, the model depth, parameter amounts, data augmentation and normalization, or training strategies can also cause performance differences. Most teams also report the performance gains from model or inputs ensemble strategies. \\

\textbf{T-OLED v.s. P-OLED}
According to the experiment results, T-OLED demonstrates an easier task than P-OLED. The T-OLED restoration problem itself resembles a combination of deblurring and denonising tasks. However, imaging through P-OLED suffers heavily from lower light transmission and color shift. Some teams like CILab IITM, Image Lab and lyl, which participated in both tracks, chose to use the same models for two tracks, while other teams tend to use different model structures. Team BigGuy explored different module options to better resolve the low-light and color issues of P-OLED inputs. Team CET$\_$CVLab addresses the information loss issues of downsampling from P-OLED inputs by using a wavelet-based decomposition. Team San Jose Earthquakes added an additional color correction stage for P-OLED. \\

\textbf{Inference Efficiency}
We did not rank the methods by the inference time due to device and platform difference of different methods. Most methods run about 1 to 3 seconds per image of size $1024\times2048$ on GPUs. Without further optimization, these models may not be easily applied in mobile devices or laptops for real-time inference of live streams or videos. However, it is still feasible to restore degraded images or videos in an offline way. Team San Jose Earthquakes run a longer inference time since their method involves an additional ADMM optimization process. Team CET$\_$CVLab claimed to achieve 0.044s inference time on a better GPU, which makes the method both outperformed and high-efficient. 

\section{Conclusions}
We summarized and presented the methods and results of the first image restoration challenge on Under-Display Camera (UDC). The testing results represent state-of-the-art performance on Under-Display Camera Imaging. Participants extensively explored state-of-the-art deep-learning based architectures conventionally used for image restoration. Some additional designs like shade and color correction are also proven beneficial to be adaptive to the through-display imaging tasks. However, the results are specifically limited to two display types, P-OLED and T-OLED, and a single camera. This further suggests the need for exploring more display types and hardware set-ups so the model can be generalized.\\
{\bf Acknowledgment}\\
We thank the UDC2020 challenge and RLQ workshop Sponsor: Microsoft Applied Science Group.
\clearpage
%
%
\bibliographystyle{splncs04}
\bibliography{egbib}
\end{document}